\documentclass{elsart3p}
\usepackage{graphicx}

\usepackage{amssymb}

\begin{document}

\begin{frontmatter}



\title{An Inhomogeneous Josephson Phase in Thin-Film and High-$T_c$ Superconductors}
%


\author{Y.Imry$^a$}, \author{M. Strongin$^b$} and \author{C. C. Homes$^b$}

\address[a]{Department of Condensed-Matter Physics, the Weizmann Institute of
Science, Rehovot, Israel, 76100}

\address[b]{Condensed Matter Physics and Materials Science Department, Brookhaven
National Laboratory, Upton, NY, 11973 USA}
%
%
\begin{abstract}
In many cases inhomogeneities are known to exist near the metal (or
superconductor)-insulator transition, as follows from well-known domain-wall
arguments.  If the conducting regions are large enough (i.e. when the $T = 0$
superconducting gap is much larger than the single-electron level spacing), and
if they have superconducting correlations, it becomes energetically favorable
for the system to go into a Josephson-coupled zero-resistance state before
(i.e. at higher resistance than) becoming a ``real'' metal.  We show that this
is plausible by a simple comparison of the relevant coupling constants. For
small grains in the above sense, the  {\it electronic} grain structure is
washed out by delocalization and thus becomes irrelevant. When the proposed
``Josephson state'' is quenched by a magnetic field, an insulating, rather then
a metallic, state should appear. This has been shown \cite{Tu} to be consistent
with the existing data on oxide materials as well as ultra-thin films. We
discuss the Uemura correlations versus the Homes law, and derive the former for
the large-grain Josephson array (inhomogenous superconductor) model. The
small-grain case behaves like a dirty homogenous metal. It should obey the
Homes law provided that the system is in the dirty supeconductivity limit. A
speculation why that is typically the case for $d$-wave superconductors is
presented.
\end{abstract}

\begin{keyword}
Inhomogenous Superconductivity, Josephson Phase, (Super)Conductor--Insulator
Transition, $n_s$--$T_c$ Correlations.

\PACS   74.40.+k, 74.50.+r, 74.80.-g
\end{keyword}
\end{frontmatter}

%
%
\section{Introduction}
\vspace*{-0.2cm}
The intriguing connection between underdoped high-$T_c$ superconductivity and
the properties of disordered and granular superconductors \cite{Tu,1,2} has
been discussed since the discovery of these superconductors. We argue in this
paper that near the superconductor-insulator (S-I) transition, inhomogeneities
may lead to a zero-resistance Josephson-coupled state, which exists both in
high-temperature superconductors and ``usual'' superconductors even though the
interactions \cite{3} causing the superconducting state may indeed be very
different. Here we review and strengthen the arguments of Ref.~\cite{Tu} and
discuss their relevance to the intriguing $n_s$ -- $T_c$ correlations
\cite{Uemura,Homes,Homes2,Lemberger}.

The underlying principle is that disorder implies inhomogeneities on {\it some}
length-scales, as was first argued, in this context, by Kowal and Ovadyahu
\cite{4}. These scales depend on the nature and strength of the disorder. This
picture is supported by numerous experiments \cite{4,5} and may be related  to
domain formation by random-field-type impurities \cite{6}, see also \cite{7}.
For example, if the Mott-type metal-insulator transition were in fact first
order, as originally argued by Mott, then the arguments of Ref. \cite{6} would
imply ``domain'' formation in effectively 2D systems even for the weakest
strength of the impurities! Finite-strength impurities will generically lead to
domain formation in most situations, except very close to an appropriate
second-order transition in cases where the correlation length diverges strongly
enough \cite{7}. Experiments considering the effect of inhomogeneities brought
about by fluctuations in the local electron density or concentration gradients
already exist in the literature \cite{5}. On the theoretical side, the
importance of inhomogeneities has been highlighted by Emery, Kivelson and
co-workers \cite{8},and by Dagotto and co-workers \cite{9}. Ghosal et al.
\cite{10} have considered a model based on the Bogoliubov-de Gennes equations,
of how ``homogeneous'' disorder introduces an inhomogeneous pairing amplitude
in ultra-thin films. We would like to add to these interesting models that in
the non-superconducting state, the phases of these domains are not locked and
therefore the phase fluctuations should average the local pairing amplitude,
$\Delta$, to zero (however, $<|\Delta|^{2}>\neq 0)$. Refs.~\cite{11} and
\cite{12} discuss Bose-Hubbard models and cite earlier theoretical references
related to disordered systems. In recent work on ultra-thin films (\cite{13},
see also Ref.~\cite{14}), it was argued that in an inhomogeneous medium it is
possible for a Josephson-coupled superconducting state to be more stable at or
near the S-I transition boundary (more disorder/less carrier density) than the
metallic state (which is defined here as being on the metallic side of the
percolation/localization transition). This general problem was treated some
years ago in Ref.~\cite{14} using considerations based on the Thouless
\cite{15} arguments for the onset of localization in 1D and handling the
Coulomb effects in the spirit of the phenomenological arguments of Abeles and
Sheng \cite{16} and Kawabata \cite{17}. This will be reviewed in section
\ref{scales} below. A simple case where this clearly works is an array of
Josephson coupled clusters with an energy gap that is larger \cite{14,18} than
the energy level spacing in the cluster (see below). In this paper we are
interested in extending these ideas to give some insight into weak
superconductivity in inhomogeneous systems, and thus whether we can understand
data in films as well as in underdoped high-$T_c$ superconductors. In
particular, we will present some insights on the Uemura correlation
\cite{Uemura} and the ``Homes law'' \cite{Homes,Homes2} (see also
\cite{Lemberger}).

\section{Scales for inhomogenous and granular systems}
\label{scales}
\vspace*{-0.2cm}
Here we briefly describe the simple argument which indicates that in an
inhomogeneous system there may be a regime in which an inhomogenous
Josephson-coupled state occurs before the metallic state, as the sample
resistance decreases from a resistance characteristic of the insulating state
to that of a conducting one.  This is done by either increasing the doping in
the high-$T_c$ case, or changing the thickness in the ultra thin film case.

Without interactions, the Thouless picture of localization in one dimension can
be generalized to analyze the electronic couplings between ``metallic regions''
\cite{15} in an inhomogeneous system (which can consist of grains or doped
regions with high conductivity) in any dimension \cite{19,20}. The intergrain
coupling energy is given by $\hbar/\tau_L = V_L$ where $\tau_L$ is the lifetime
for an electron in one of the conducting regions, of linear size $L$, to go
into the next one. The conductance between ``grains'' can be related to the
ratio of this coupling energy to the energy level spacing in the grains and is
written as a dimensionless conductance, $g_L=V_L/w_L = 2\hbar/e^2R_L$, where
$w_L$ is the characteristic energy-level spacing in the small metallic regions.
When the typical intergrain resistance, $R_L > 2\hbar/e^2$ then the
noninteracting system becomes localized.

We now introduce the simplified Coulomb interaction, parametrized by a single
capacitive energy, see below. We start with the analysis of Abeles and Sheng
\cite{16} to estimate the resistance between isolated grains where the coupling
energy overcomes the intergranular Coulomb energy of $E_{coul} = e^2/2C_L$. By
approximating $\hbar/\tau_L$ as $\hbar/R_LC_L$, and setting this equal to
$e^2/2C_L$, one gets the same value as before, of $R_L \simeq 2\hbar/e^2$ for
the resistance below which the ``intergranular'' coupling is greater than the
Coulomb repulsion (where $R_L$ is the tunneling resistance between grains and
$C_L$ is the mutual capacity of the two grains). Thus in this case a system
with Coulomb interactions will also be metallic once $R_L < 2\hbar/e^2$.
Clearly, these two approaches are not unrelated. A physical argument relating
them might be based on the fact that once the single-electron eigenfunctions
are delocalized and spill over from the grain, the Coulomb blockade picture
with quantized charge on the grain becomes meaningless. Evidently, this
argument is certainly valid when the Coulomb energy is weak, $e^2/2C_L < w_L$,
and it is treated as a perturbation on the noninteracting picture. The argument
may also hold for strong interactions, $e^2/2C_L > w_L$, provided that {\it the
actual value for $R_L$, which may be strongly renormalized by the interaction,
is used}. In Ref.~\cite{14} the noninteracting picture was generalized,
following Ref.~\cite{17}, to include the effect of strong Coulomb interactions
(i.e. $e^2/2C_L > w_L$), which of course is typically crucial due to the
marginal screening and the charging energy when electrons move between
conducting regions. Here, to get metallic behavior, the intergrain transfer
energy should overweigh the Coulomb energy \cite{14,17} (see also
\cite{19,20}):
\begin{equation}
  zV_L=2\hbar w_Lz/e^2R_L>e^ 2/2C_L,
  \label{1}
\end{equation}
where $z$ is the coordination number, related to the typical number of nearest
neighbors, which appears in the mean-field theory for the transition.

The condition for superconductivity is, however, that the Josephson energy,
given by the standard expression $E_J = \pi \hbar \Delta (0)/4e^2 R_L,$ be
larger than the Coulomb energy, or, putting again the factor $z$ for a medium
composed of grains we replace $E_J$ by $z E_J \sim zV_L \Delta(0)/w_L$,
\begin{equation}
 z \pi \hbar \Delta(0)/4e^2 R_L>e^2/2C_L.
 \label{2}
\end{equation}
Here $\Delta(0)$ is the gap at $T=0$. In other words, the pair transfer matrix
element $E_J$ replaces here the single-electron coupling energy $V_L$. For this
approximate argument at low temperatures, there is no need to put in the
temperature dependence of the gap. The interesting consequence is that there
clearly exists an unusual regime where $E_{coul}$ can be greater than $zV_L$,
but less than $E_J$, as long as $\Delta(0)/w_L$ is greater than 1. So for
grains that are ``large'' in the sense \cite{18} that $\Delta(0)/w_L\gg 1$,
superconductivity is easier to achieve than normal conductivity
\cite{Tu,14,19}. This argument is only meant to show that if there are
intrinsic inhomogeneities and the system has superconducting regions, then {\it
it is possible to have a Josephson state before having a metallic one.}

A possible phase diagram, for $\Delta(0) \gg w_L$, is described in Fig.~1. It
can be seen that at low temperatures as the conductivity increases, (by
increasing the thickness in the case of films and increasing the doping in
underdoped high $T_c$'s) one first goes from the insulating phase into the
Josephson phase (line A-B) and finally into the true metallic/superconducting
phase where  the respective single-particle ``wavefunctions'' become
delocalized \cite{Tu,14,25}. (Note that line C-B and its continuation to higher
temperatures, eventually becomes a smooth crossover rather then a sharp
transition. We do not know whether the change from the Josephson phase to the
usual superconductor is affected by a real transition). This is consistent with
the above argument showing that if there are superconducting correlations in
the insulating regime, then the quantum transition to a Josephson state can
occur [for large ``grains'' where  $\Delta(0)/w_L >1$] before the
percolation-delocalization transition.

%
%
\begin{figure}[t]
\vspace*{-0.5cm}%
\centerline{\includegraphics[width=8cm,clip]{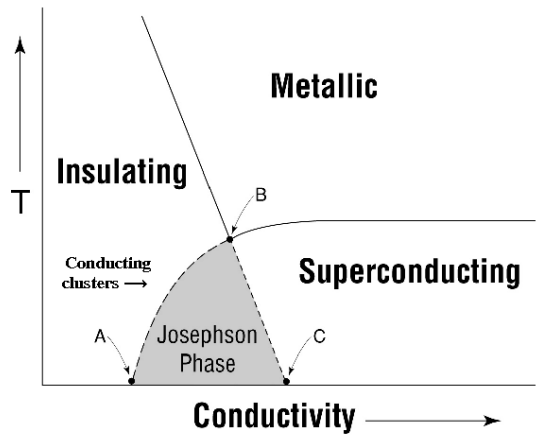}}%
\vspace*{-0.2cm}%
\caption{The schematic ``phase diagram'' of an inhomogeneous ``large-grain''
[$\Delta(0)/w_L \gg 1$] superconductor in the Temperature/Conductivity plane.
The conductivity is used as a measure of disorder and increases with doping or
film thickness. The A-B line is the boundary between the insulating phase and
the Josephson coupled state where there are isolated superconducting regions
that are Josephson coupled.  The B-C line is where the system goes into a
delocalized bulk superconducting phase .  In this region to the right of line
B-C we would expect normal metallic conduction when superconductivity is
quenched by a magnetic field. In the Josephson phase a logarithmic behavior in
the resistivity seems common when superconductivity is quenched by a field. The
note in the figure indicates a region to the left of line A-B where there exist
disconnected metallic regions with nonzero $<|\Delta|^{2}>$ at low
temperatures. Likewise, disconnected insulating regions may occur to the right
of the line A-B. }%
\label{Fig1}
\end{figure}

What happens for small grains $\Delta(0)/w_L \ll 1$? Here, by decreasing
disorder, one goes from the insulator first into the metallic, delocalized
phase. In this metallic phase the single-electron wavefunctions are no longer
localized in the grains. The granular inhomogeneity is thus not effective for
the electrons! Therefore, by decreasing the disorder further, or increasing the
interactions responsible for superconductivity, {\it the system will go into
the superconducting state as a continuous (not a granular!) system}.
%
%

Does experimental evidence exists for our conjectures?  Some of this evidence,
in both ultra thin films and underdoped high-temperature superconductors, has
been discussed in Ref.~\cite{Tu}. In the next section, \ref{Uemura-Homes}, we
review the correlations (discovered experimentally) between the superfluid
density $n_s$ and the transition temperature $T_c$ and how they can be easily
understood from our picture. Before doing that, we remark that the insulating
state with intragrain superconducting correlations and $<|\Delta|^{2}> \neq 0$,
but without intergrain phase locking, provides a simple example where a local
``pseudogap'' may exist above $T_c$. This pseudogap will have, with respect to
the local ``crystal'' axes, the same symmetry and nature as the superconducting
gap below $T_c$. This agrees with recent angle-resolved photoemission studies
\cite{27}.

%
%
\section{The $n_s$ -- $T_c$ correlations}
\label{Uemura-Homes}
\vspace*{-0.2cm}
Here we consider the question of the universal correlations reported
experimentally between the low-temperature superfluid density, $n_s$ and the
transition temperature $T_c$. Three  such correlations have been reported, for
underdoped high-$T_c$ superconductors and in some cases for usual ``low-$T_c$''
ones. Two of them are different from each other, while the third may be related
to the second (see below). It is of great interest to understand the physics
behind such correlations and what are their respective ranges of validity.

In 1989, Uemura et al. \cite{Uemura} reported the proportionality of $n_s/m^*$
(or $\lambda^{-2}$, where $\lambda$ is  the penetration length and $m^*$ the
carrier mass, which is of the order of $5m$ for the considered materials) to
$T_c$; $n_s$ was determined from the muon spin relaxation rate for four high
$T_c$ families with varying doping level (carrier density). The coefficient in
the linear relationship is such that a carrier density of $2\times 10^{21}/{\rm
cm}^3$ corresponds to $T_c\simeq 25$~K.

In 2004, Homes et al. \cite{Homes} reported a different correlation: $N_c
\simeq 4.4\,\sigma_{dc} T_c$, where $N_c = n_s/8$ is the spectral weight,
determined by optical measurements, associated with the superconducting
condensate and $\sigma_{dc}$ is the normal-state dc conductivity near $T_c$.
Nine different high-$T_c$ material families with varying doping (including
optimal and beyond) were examined, as well as the usual superconductors Pb and
Nb. This result has been interpreted \cite{Homes2} in terms of the conventional
decrease of $n_s$ proportional to $\ell/ \xi_0 \propto T_c \tau $ in the dirty
limit of BCS superconductors, where $\tau$ and $\ell$ are the mean free time
and scattering length and $\xi_0$ the zero-temperature BCS coherence length
($\xi_0 \propto v_F/T_c$). The questions of why these materials are in the
dirty limit, when $T_c$ is so high and to what extent can the BCS-type
relationships be used for high-$T_c$ materials (in spite of current theoretical
beliefs) were left open. Clearly, the $d$-wave nature of these superconductors
might play an important role here.

Finally, in 2005 Zuev et al. \cite{Lemberger} reported a linear relationship
between $n_s$ and $T_c^{^\chi}$, where $\chi = {2.3 \pm 0.4}$. They pointed out
that with the empirical  proportionality of $T_c$ to $\sigma_{dc}$
(theoretically justified in a classical Josephson-coupled superconductor
\cite{14}), the value of $2$ for $\chi$ makes their result consistent with the
one by Homes et al. \cite{Homes}. Obviously $\chi = 2$ is well within the
experimentally determined range.

We shall now present a derivation of the $n_s \propto T_c$ relation for a
classical ordered Josephson array under the assumption that the size, $L$, of
each superconducting unit is $\ll\lambda$. This can be taken as a model for a
granular superconductor as long as the effect of intergrain disorder, which
certainly exists in real cases, is not dominant.

Consider for simplicity a regular infinite 2D array of square superconducting
grains of linear size $L$ and thickness $d$, connected by flat Josephson
junctions with Josephson current amplitudes $I_J$ and Josephson energies $E_J =
\hbar I_J / 2e$. The generalization to a 3D array is straightforward. We obtain
the linear response to a small magnetic field $B$ perpendicular to the array.
For $\lambda \gg L$ the field $B$ is uniform over each grain. $B$ is derived
from a vector potential $\vec A = (By, 0, 0)$. Note that $div A = 0$ as
required for the London gauge. Thus the London equation takes the form
\begin{equation}
  j_s = -\frac{n_s e^2}{m^* c} A.
  \label{London}
\end{equation}

Due to the flux, the phase difference between two superconducting blocks that
are nearest neighbors in the $x$ direction, increases with $y$ in the manner
\begin{equation}
  \phi(y) \simeq -2eByL/\hbar = -2eL A_x(y)/\hbar.
  \label{phase}
\end{equation}
For small B, this leads to a Josephson current density
\begin{equation}
  j_{s,x}(y) = -2eI_J A_x(y)/\hbar.
  \label{current}
\end{equation}
Comparing with the London equation (\ref{London}), we find the Uemura-type
relation:
\begin{equation}
  n_{s} = \frac{4m^*}{d\hbar^2 \zeta} T_c,
  \label{ns}
\end{equation}
where the constant of order unity $\zeta$ is defined via $T_c = \zeta E_J$, and
we have used units in which $k_B = 1$ throughout. For $m^{*} = 5m$, $\zeta =
1$, $d = 5$~\AA\ and $n_s = 2\times 10^{21}\,{\rm cm}^{-3}$ we obtain $T_c
\approx 35$~K. Thus, the coefficient in eq.~\ref{ns} agrees within a factor of
two with the Uemura one, for reasonable parameters of the 2D layer.
Eq.~\ref{ns} is just the relation between $n_s$ and the order-parameter phase
stiffness for the $x-y$ model.

When the Uemura correlation was first reported, the proportionality of $T_c$ to
the 2D electron density was taken to indicate the purely electronic origin of
high-$T_c$ superconductivity. Our simple derivation above, proves that that
logic is not infallible. The Josephson array can model any appropriately
inhomogenous superconductor, including ordinary low-$T_c$ ones, and it does
yield the Uemura correlation.

The Uemura correlation should thus be valid for the large grains,
$\Delta(0)/w_L > 1$ case, where the inhomogenous Josephson phase is the
relevant one. In the small grain case $\Delta(0)/w_L < 1$, superconductivity is
established in a homogenous, strongly disordered, conductor. Close to the
metal-insulator transition the mean free path $\ell$ is of a small microscopic
magnitude and it makes sense that the superconductor should be in the dirty
limit ($\ell \ll \xi_{0}$). This implies \cite{Homes2} that the Homes law
\cite{Homes,Homes2} (or the one reported by Zuev et al. \cite{Lemberger})
should then yield the valid correlation between $n_s$ and $T_c$. The case of
high-$T_c$ materials is further complicated due to the anisotropic gap and
correlation length. The question of when can such a superconductor be regarded
as dirty is interesting and nontrivial. Its full analysis is beyond the scope
of this paper. We speculate in the next section that the fact that in the nodal
regions $\Delta(0)\tau \ll 1$ even for weak disorder, may well be relevant.

%
%
\section{Some experiments on inhomogenous superconductors and thoughts
on High $T_c$.} \label{exp}
\vspace*{-0.2cm}
In this short section we would like to discuss how the inhomogeneities we have
previously mentioned can be consistent with the correlations found by Homes et
al. \cite{Homes,Homes2}.  These issues involving inhomogeneities occur not only
in the high-$T_c$ superconductors, but also in some other granular systems. For
example we briefly mention that both underdoped high-$T_c$ superconductors
\cite{28} and other systems such as NbN \cite{29} show a logarithmic dependence
with temperature for the resistance in a magnetic field. Beloborodov et al.
\cite{30} have mentioned that this logarithmic behavior can occur in granular
systems and high-$T_c$ superconductors. Furthermore, tunneling microscopy
(although not completely understood) also shows evidence that high-$T_c$
superconductors are not physically homogeneous \cite{5}. We have previously
mentioned that there are two limits for inhomogenieties in superconductors.
There is the Josephson phase where $\Delta(0)/w_L \gg 1$, and the small grain
case where $\Delta(0)/w_L \ll 1$ and the various regions of the film are
connected.

A summary of the nature of the films that satisfy these conditions has been
given some time ago \cite{Tu,14}. The small grain case is relatively simple and
will not be further discussed in this paper. In the high-$T_c$ superconductors
the situation is more complex and in the rest of this section we would like to
deal with Homes' law \cite{Homes,Homes2} and why it works there. Tunneling
microscopy \cite{5} seems to indicate that high-temperature superconductors are
consistent with the condition that $\Delta(0)/w_L > 1$. We have already
indicated that Homes' law seems consistent with a dirty superconductor and in
fact it works for dirty conventional metallic films as well as high-$T_c$
superconductors. The question is why this works for high-temperature
superconductors, which microscopy shows are granular in nature. Granular
systems where $\Delta(0)/w_L \gg 1$ can indeed be superconducting, but they do
not necessarily act like dirty-limit superconductors in the Ginzburg-Landau
theory, unless the coherence length is long. It is possible that the $d$-wave
nature in high temperature superconductors provides a way for this to happen.
In the nodal region the superconducting gap is small and the coherence length
may be large. These considerations are discussed in an interesting paper by
Joglekar et al. \cite{Jog}. This situation, where there is a large coherence
length (in the nodal region) compared to the mean free path, may bring the
system back to the dirty limit and this could serve as a possible way to
understand Homes' law in high-$T_c$ superconductors.

%
%
\section{Conclusions}
\vspace*{-0.2cm}
 We proposed the  picture of spontaneously formed conducting
domains which form a Josephson phase at low temperatures, as a general
description for some disordered systems near the superconductor-insulator
transition, especially in the effectively 2D case \cite{6}. As far as we know,
there is really no experimental evidence for a uniform state at the S-I
transition. In this regime where large superconducting regions first appear (to
the left of line A-B in Fig.~1) they are initially decoupled (this region which
is analogous to the pseudo-gap state is not shown).  As line A-B is approached,
Josephson coupling produces phase alignment of the order parameter of different
superconducting regions, and this happens before the percolation-delocalization
transition to the metallic state, along line C-B. We discussed the various
$n_s$ -- $T_c$ correlations and showed how the Uemura correlation naturally
arises for the Josephson array superconductor. The Homes correlations follow
for a dirty superconductor and we speculated how this can arise in the $d$-wave
case.

%
%
{\bf Acknowledgements} We thank Amnon Aharony, Joe Bhaseen, John Chalker, Sasa
Dordevic, Alexander Finkelstein,  S. A. Kivelson, Robert Konik, Zvi Ovadyahu,
Alexei Tsvelik, John Tranquada,  Jim Valles, Matthias Vojta, and Peter
W\"olfle, for helpful discussions. We are grateful to T. Valla for many
discussions and for sharing his insights about the issues of $d$-wave
superconductivity, the nodal regions and their possible implications for Homes'
law. We are also grateful to J. Tu for collaborations and discussions in some
of the earlier work. Work at Brookhaven was supported by the Department of
Energy (DOE) under Contract No. DE - AC02-98CH10866. Work of Y. I. was
supported by a Center of Excellence of the Israel Science Foundation (ISF,
grant No. 1566/04), and by the German Federal Ministry of Education and
Research (BMBF) within the framework of the German-Israeli project cooperation
(DIP).




\begin{thebibliography}{00}
\bibitem{25} We emphasize that, in the case of ultra thin films,  $\Delta(0)/w_L$
is large as long as the extent of the inhomogeneous regions contains a few
thousand electrons and Cooper pairs can form.  Hence, in this case, the extent
must be at least of the order of 10 nm.
\end{thebibliography}
\end{document}